\documentclass[a4paper,12pt]{article}
\usepackage{amsmath}
\usepackage{graphicx}
\usepackage[english]{babel}
\pdfpagewidth
\pdfpageheight  
\title{\textbf{Twisted Skyrmion String\footnote{Presented at CTPNP 2014 and submitted to Journal of Physics: Conference Series.}}}
\author{Miftachul Hadi$^{\varepsilon,1,2,4,5}$,~Malcolm Anderson$^{1}$,~Andri Husein$^{3}$\\\\
$^1$Department of Mathematics, Universiti Brunei Darussalam\\ 
Jalan Tungku Link BE1410, Gadong, Negara Brunei Darussalam
\\\\
$^2$Physics Research Centre, Indonesian Insitute of Sciences (LIPI)\\
Kompleks Puspiptek, Serpong, Tangerang 15314, Indonesia
\\\\
$^3$Department of Physics, University of Sebelas Maret\\
Jalan Ir. Sutami 36 A, Surakarta 57126, Indonesia
\\\\
$^4$Department of Physics, School of Natural Sciences\\
Ulsan National Institute of Science and Technology (UNIST)\\
50, UNIST-gil, Eonyang-eup, Ulju-gun, Ulsan, South Korea
\\\\
$^5$Institute of Modern Physics, Chinese Academy of Sciences\\
509 Nanchang Rd., Lanzhou  730000, China\\\\
$^\varepsilon$E-mail: itpm.id@gmail.com}
\date{}
\begin{document}
\maketitle
\pagenumbering{arabic}
\begin{abstract}
We study nonlinear sigma model, especially Skyrme model without twist and Skyrme model with twist: twisted Skyrmion string. Twist term, $mkz$, is indicated in vortex solution. Necessary condition for stability of vortex solution has consequence that energy of vortex is minimum and scale-free (vortex solution is neutrally stable to changes in scale). We find numerically that the value of vortex minimum energy per unit length for twisted Skyrmion string is $20.37\times 10^{60}~\text{eV/m}$.
\end{abstract}

\section{Introduction to Nonlinear Sigma Model}
Nonlinear sigma model is a $n$-component scalar field theory where the field defines a mapping from space-time to a target manifold. A mapping here means a function from space-time to the target space \cite{Zakrzewski}. 

By a nonlinear sigma model, we mean a field theory with the following properties \cite{hans02}:
\begin{itemize}
\item[(1)] The fields, $\phi(x)$, of the model are subjected to nonlinear constraints for all points $x\in\mathcal{M}_0$, where $\mathcal{M}_0$ is source (base) manifold, i.e. the spatial submanifold of the (2+1) or (3+1)-dimensional space-time manifold.
\item[(2)] The constraints and the Lagrangian density are invariant under the action of a global (space independent) symmetry group, $G$, on $\phi(x)$.
\end{itemize}

The Lagrangian density of a free (without potential) nonlinear sigma model on a Minkowski background space-time is defined as \cite{chen}
\begin{equation}\label{1}
\mathcal{L}=\frac{1}{2\lambda^2}~g^{ab}(\phi)~\eta^{\mu\nu}~\partial_\mu\phi_a~\partial_\nu\phi_b
\end{equation}
where $g^{ab}(\phi)$ is field metric, $\eta^{\mu\nu}$ is the Minkowski metric tensor, $\lambda$ is a scaling constant with dimensions of (length/energy)$^{1/2}$ and here $\phi$ is field. The nonlinearity is manifest in the field metric, $g_{ab}(\phi)$.  

A special case of the nonlinear sigma model occurs when the target manifold is the unit sphere $S^2$ in $R^3$, i.e. $g^{ab}(\phi)=\delta^{ab}$. In case of $a=b$ then $\delta^{ab}=1$, where $\delta^{ab}$ is Kronecker delta. The Lagrangian density (\ref{1}) then becomes
\begin{equation}\label{2}
\mathcal{L}=\frac{1}{2\lambda^2}~\eta^{\mu\nu}~\partial_\mu\phi
~.~\partial_\nu\phi
\end{equation}
where the dot (.) denotes the standard inner product on $R^3$, and the image of $\phi$ is $S^2$.
Simple representation of $\phi$ (in case of time-dependent) is
\begin{equation}\label{3}
\phi=
\begin{pmatrix}
\sin f(t,x)~\sin g(t,x) \\
\sin f(t,x)~\cos g(t,x) \\
\cos f(t,x)
\end{pmatrix}
\end{equation}
where $f$ and $g$ are scalar functions on the background space-time, with Minkowski coordinates $x^\mu=(t,x)$.

Substitute (\ref{3}) into (\ref{2}), then Lagrange density (\ref{2}) becomes
\begin{equation}\label{4}
\mathcal{L}=\frac{1}{2\lambda^2}(\eta^{\mu\nu}~\partial_\mu f~\partial_\nu f+[\sin^2f]~\eta^{\mu\nu}~\partial_\mu g~\partial_\nu g)
\end{equation}
Associated Euler-Lagrange equations from $\mathcal{L}$ (\ref{4}) are
\begin{eqnarray}\label{5}
\eta^{\mu\nu}~\partial_\mu\partial_\nu f-(\sin f~\cos f)~\eta^{\mu\nu}~\partial_\mu g~\partial_\nu g=0
\end{eqnarray}
\begin{eqnarray}\label{6}
\eta^{\mu\nu}~\partial_\mu\partial_\nu g+2(\cot f)~\eta^{\mu\nu}~\partial_\mu f~\partial_\nu g=0
\end{eqnarray}

\section{$O(N)$ Nonlinear Sigma Model}
The simplest example of nonlinear sigma models is the $O(N)$ nonlinear sigma model which consist of $N$-real scalar fields, $\phi^A,~A=1,..,N$, having the Lagrangian density \cite{hans02}
\begin{equation}\label{7}
\mathcal{L}=\frac{1}{2}~g^{\mu\nu}~\frac{\partial\phi^A}{\partial x^\mu}~\frac{\partial\phi^A}{\partial x^\nu}
\end{equation}
where the scalar fields, $\phi^A$, satisfy the constraint
\begin{equation}\label{8}
\phi^A\phi^A=1.
\end{equation}
The Lagrangian density (\ref{7}) is obviously invariant under the global (space independent) orthogonal transformations $O(N)$, i.e. the group of $N$-dimensional rotations \cite{hans02}
\begin{equation}\label{9}
\phi^A\rightarrow\phi'^A=O^A_B~\phi^B.
\end{equation}

One of the most interesting examples of $O(N)$ nonlinear sigma models due to its topological properties, is the $O(3)$ nonlinear sigma models in 1+1 dimensions, with the Lagrangian density \cite{wiki1}
\begin{equation}\label{10}
\mathcal L= \frac{1}{2}~\partial^\mu \phi \cdot\partial_\mu \phi 
\end{equation}
where $\phi=(\phi_1,\phi_2,\phi_3)$, due to $N=3$, with the constraint $\phi\cdot\phi=1$ and $\mu=1,2$. 

\section{Soliton Solution}
Two solutions to these equations (\ref{5}), (\ref{6}), are 
\begin{itemize}
\item[(i)] A monopole solution, which has
\begin{equation}\label{11}
\phi=\hat{\textbf{r}}=
\begin{pmatrix}
x/\rho\\
y/\rho\\
z/\rho\\
\end{pmatrix}
\end{equation}
where $\rho=(x^2+y^2+z^2)^{1/2}$ is the spherical radius.
\item[(ii)] A vortex solution, which is found by imposing the "hedgehog" ansatz
\begin{equation}\label{12}
\phi=\begin{pmatrix}
\sin f(r)~\sin (n\theta-\chi)\\
\sin f(r)~\cos (n\theta-\chi)\\
\cos f(r)
\end{pmatrix}
\end{equation}
where $\theta=\arctan (x/y)$, $n$ is a positive integer, and $\chi$ is a constant phase factor.
\end{itemize}
A vortex is a stable time-independent solution to a set of classical field equations that has finite energy in two spatial dimensions; it is a two dimensional soliton. In three spatial dimensions, a vortex becomes a string, a classical solution with finite energy per unit length \cite{preskill}. Solutions of finite energy, satisfying the appropriate boundary conditions, are candidate soliton solutions \cite{manton}.

Recall that there is in fact a family of vortex solutions
\begin{equation}\label{13}
\sin f=\frac{2K^{1/2}r^n}{1+Kr^{2n}}
\end{equation}
or
\begin{equation}\label{14}
\cos f=\frac{Kr^{2n}-1}{Kr^{2n}+1}
\end{equation}
For each value of $K$ where $K$ is positive constant, there is a different vortex solution.

But, the mass per unit length 
\begin{equation}\label{15}
\mu = -\frac{4\pi n}{\lambda^2}
\end{equation}
does not depend on $K$. (We use the same notation for energy per unit length and mass per unit length, due to equivalence of energy-mass as $E=mc^2$. Here, we take $c=1$).

This means that the vortex solutions are what is called neutrally stable to changes in scale. As $K$ change, the scale of the vortex changes, but the mass per unit length, $\mu$, does not. Note that because of eq.(\ref{15}), there is a preferred winding number, when $n$ is a small as possible: $n=1$. It means that for the vortex solution, the topological charge is just the winding number, $n$. 

It can be shown that the topological charge is conserved, no matter what solution $\phi$ we have. So, topological charge is a constant, no matter what nonlinear sigma model we use. So long as
\begin{eqnarray}\label{16}
\phi=
\begin{pmatrix}
\sin f~\sin g\\
\sin f~\cos g\\
\cos f
\end{pmatrix}.
\end{eqnarray}

Let us find $f$ by solving the two equations of motion (\ref{5}), (\ref{6}). The function $f$ satisfies the equation 
\begin{equation}\label{17}
r\frac{d^2f}{dr^2}+\frac{df}{dr}-\frac{n^2}{r}~\sin f~\cos f=0
\end{equation}
and that the solution satisfying the boundary conditions
\begin{equation}\label{18}
f(0)=\pi
\end{equation}
and 
\begin{equation}\label{19}
\qquad \lim_{r\to\infty}f(r) = 0
\end{equation}
is
\begin{equation}\label{20}
\cos f=\frac{Kr^{2n}-1}{Kr^{2n}+1}
\end{equation}
This is the vortex solution.

The energy density of a static (time-independent) field with Lagrangian density $\mathcal{L}$ (\ref{4}) is
\begin{eqnarray}\label{21}
E
&=&-\mathcal{L}\nonumber\\
&=&-\frac{1}{2\lambda^2}\left[\eta^{\mu\nu}~\partial_\mu f~\partial_\nu f+(\sin^2 f)~\eta^{\mu\nu}~\partial_\mu g~\partial_\nu g\right].
\end{eqnarray}
The energy density of the monopole solution is
\begin{equation}\label{22}
E=\frac{1}{\lambda^2\rho^2}
\end{equation}
and that the energy density of the vortex solutions is
\begin{eqnarray}\label{23}
E
&=&\frac{4Kn^2}{\lambda^2}\frac{r^{2n-2}}{(Kr^{2n}+1)^2}.
\end{eqnarray}
Then the total energy
\begin{equation}\label{24}
E=\int\int\int E~ dx~dy~dz,
\end{equation}
of the monopole solution is infinite. But, that the energy per unit length of the vortex solutions
\begin{equation}\label{25}
\mu=\int\int E~dx~dy=\frac{4\pi n}{\lambda^2}
\end{equation}
is finite, and does not depend on the value of $K$. 

This last fact means that the vortex solutions in the nonlinear sigma models have no preferred scale. A small value of $K$ corresponds to a more extended vortex solution, and a larger value of $K$ corresponds to a more compact vortex solution, as we can see by plotting $f$ (or $E$) for different values of $K$ and a fixed value of $n$ (say, $n=1$).

But, the value of the energy per unit length, $\mu$, is the same for all these solutions, and so there is no natural size for the vortex solutions. It is for this reason that a Skyrme term is added to the Lagrangian density \cite{malcolm}. 

\section{Skyrmion without Twist: Skyrme Model}
We need to add Skyrme term to the Lagrangian to stabilize the vortex (which is neutrally stable to cylindrically symmetric perturbations). Original sigma model Lagrangian (in unit sphere) is
\begin{equation}\label{26}
\mathcal{L}_1=\frac{1}{2\lambda^2}~\eta^{\mu\nu}~\partial_\mu\phi~.~\partial_\nu\phi
\end{equation}
Adding a Skyrme term to eq.(\ref{26}), then eq.(\ref{26}) becomes
\begin{eqnarray}\label{27}
\mathcal{L}_2
&=&\frac{1}{2\lambda^2}~\eta^{\mu\nu}~\partial_\mu\phi~.~\partial_\nu\phi-\underbrace{K_s~\eta^{\kappa\lambda}~\eta^{\mu\nu}(\partial_\kappa\phi\times\partial_\mu\phi)~.~(\partial_\lambda\phi\times\partial_\nu\phi)}_{\text{Skyrme term}}
\end{eqnarray}
Eq.(\ref{27}) can be written in other expression by substituting (\ref{16}) into (\ref{27}). We get
\begin{eqnarray}\label{28}
\mathcal{L}_2
&=&\frac{1}{2\lambda^2}\left(\eta^{\mu\nu}~\partial_\mu f~\partial_\nu f+\sin^2f~\eta^{\mu\nu}~\partial_\mu g~\partial_\nu g  \right)\nonumber\\
&&-~K_s\left[2\sin^2f~\left(\eta^{\mu\nu}~\partial_\mu f~\partial_\nu f\right)\left(\eta^{\kappa\lambda}~\partial_\kappa g~\partial_\lambda g\right)-2\sin^2f~\left(\eta^{\mu\nu}~\partial_\mu f~\partial_\nu g\right)^2\right]
\end{eqnarray}
The Skyrme term becomes the second term on the right hand side of eq.(\ref{28}). At this point, we need to write out Euler-Lagrange equations from $\mathcal{L}_2$ (\ref{28}), i.e.
\begin{eqnarray}\label{29}
\partial_\alpha\left(\frac{\partial\mathcal{L}_2}{\partial(\partial_\alpha f)}\right)-\frac{\partial\mathcal{L}_2}{\partial f}=0~~~~~~~;~~~~~~~~\partial_\alpha\left(\frac{\partial\mathcal{L}_2}{\partial(\partial_\alpha g)}\right)-\frac{\partial\mathcal{L}_2}{\partial g}=0
\end{eqnarray}

Energy can be derived from $\mathcal{L}_2$ (\ref{28}), as
\begin{eqnarray}\label{30}
E=\int\int\left\{\frac{1}{2\lambda^2}\left[\left(\frac{df}{dr}\right)^2+\frac{n^2}{r^2}~\sin^2f\right]-2K_s~\frac{n^2}{r^2}~\sin^2f~\left(\frac{df}{dr}\right)^2\right\}r~dr~d\theta
\end{eqnarray}
Let us define new variable
\begin{eqnarray}\label{31}
\overline{r}\equiv qr
\end{eqnarray}
where $q$ is a constant. Then
\begin{eqnarray}\label{32}
\frac{df}{dr}=\frac{\partial f}{\partial\overline{r}}~q
\end{eqnarray}
where $\partial\overline{r}=q~\partial r$. So, the energy (\ref{30}) can be rewritten using new variables as
\begin{eqnarray}\label{33}
E=\int\int\left\{\frac{1}{2\lambda^2}\left[\left(\frac{\partial f }{\partial\overline{r}}\right)^2+\frac{n^2}{r^2}~\sin^2f\right]-2q^2~K_s~\frac{n^2}{\overline{r}^2}~\sin^2f~\left(\frac{\partial f}{\partial\overline{r}}\right)^2\right\}\overline{r}~d\overline{r}~d\theta
\end{eqnarray}

From (\ref{33}), if we let $q\rightarrow\infty$ then the energy per unit length goes to $-\infty$. So, it is energetically favourable for the vortex to evolve , so that $q$ increases. Then,
\begin{eqnarray}\label{34}
f(r)=f\left(\frac{\overline{r}}{q}\right)
\end{eqnarray}
and as $q\rightarrow\infty$ for fixed $r$, $\overline{r}\rightarrow\infty$ and the field evaporates to infinity.
To fix this problem, we add a potential term, $K_v(1-\underline{n}.\hat{\underline{\phi}})$, to Lagrangian density $\mathcal{L}_2$. So, we have
\begin{eqnarray}\label{35}
\mathcal{L}_3=\mathcal{L}_2+K_v(1-\underline{n}.\hat{\underline{\phi}})
\end{eqnarray}
where $\underline{n}$ is a direction of $\hat{\underline{\phi}}$ at $r=\infty$ (where, $f=0$).

This $\mathcal{L}_3$ model is like Baby Skyrmion model \cite{piette} p.207, eq.(2.2). The kinetic term along with the Skyrme term are not sufficient to stabilize a baby Skyrmion, contrary to the usual Skyrme model. The kinetic term in $2+1$ dimensions enjoys (suffers from) conformal invariance and the baby Skyrmion can always reduce its energy by inflating (infinitely).
Hence, one adds the mass term which limits the size of the baby Skyrmion. The usual Skyrme term of course prohibits the collapse of the soliton \cite{gisiger}.

\section{Skyrmion with Twist: Twisted Skyrmion String}
Back to Skyrme model (\ref{28}) and refer to eq.(\ref{16}). Instead of choosing
\begin{equation}\label{36}
g=n\theta-\chi
\end{equation}
we choose
\begin{equation}\label{37}
g=n\theta+mkz
\end{equation}
where $mkz$ is twist term. Then eq.(\ref{28}) becomes
\begin{eqnarray}\label{38}
\mathcal{L}_2=\frac{1}{2\lambda^2}\left[\left(\frac{df}{dr}\right)^2+\sin^2f\left(\frac{n^2}{r^2}+m^2k^2\right)\right]-2K_s~\sin^2f\left(\frac{df}{dr}\right)^2\left(\frac{n^2}{r^2}+m^2k^2\right)
\end{eqnarray}

Twist is identified as direction of particle which rotates circularly around string (string can be imagined e.g. as a rod in $z$ axis). The direction of twist can be clock-wise or counter clock-wise. There is a different value of "pressure" in clock-wise and counter clock-wise directions. Pressure is related with energy, it means that pressure is also related with mass, due to energy-mass relation \cite{malcolm}.

Euler-Lagrange equation from $\mathcal{L}_2$ (\ref{28}) with twist term (\ref{32}), twisted Skyrmion string, is
\begin{eqnarray}\label{39}
0
&=&\frac{1}{\lambda^2}\left[\frac{d^2f}{dr^2}+\frac{1}{r}\frac{df}{dr}-\left(\frac{n^2}{r^2}+m^2k^2\right)\sin f\cos f\right]
-4\left(\frac{n^2}{r^2}+m^2k^2\right)K_s\sin^2f\left(\frac{d^2f}{dr^2}-\frac{1}{r}\frac{df}{dr}\right)\nonumber\\
&&-~4\left(\frac{n^2}{r^2}+m^2k^2\right)K_s~\sin f~\cos f~\left(\frac{df}{dr}\right)^2
\end{eqnarray}

\section{Twisted Skyrmion String: Numerical Calculation and Result}
Refer to eq.(\ref{39}), let us rewrite Euler-Lagrange equation from $\mathcal{L}_2$, eq.(\ref{28}), i.e. twisted Skyrmion string as
\begin{eqnarray}\label{40}
\frac{d^2f}{dr^2}
&=&-~\frac{(\varepsilon+\zeta~r^2)~\sin f~\cos f}{r^2+(\varepsilon+\zeta~r^2)~\sin^2f}~\left(\frac{df}{dr}\right)^2-\frac{1}{r}~\left(\frac{r^2-(\varepsilon+\zeta~r^2)~\sin^2f}{r^2+(\varepsilon+\zeta~r^2)~\sin^2f}\right)~\frac{df}{dr}\nonumber\\
&&+~\frac{n^2(1+\frac{\zeta}{\varepsilon}~r^2)~\sin f~\cos f}{r^2+(\varepsilon+\zeta~r^2)~\sin^2f}
\end{eqnarray}
Eq.(\ref{40}) can be solved numerically for different values of $\varepsilon$ and $\zeta$, starting with $f(0)=\pi$, $f'(0)=-a$ for different values of $a$. We need the numerical solution of (\ref{40}) for calculating minimum energy of vortex.

Let us use Runge-Kutta fourth order method for solving (\ref{40}). First, assume that
\begin{eqnarray}\label{41}
\frac{df}{dr}=v
\end{eqnarray}
\begin{eqnarray}\label{42}
\frac{d^2f}{dr^2}=\frac{dv}{dr}
\end{eqnarray}
Then
\begin{eqnarray}\label{43}
\frac{dv}{dr}
&=&g(r,~f,~v)\nonumber\\
&=&-\frac{(\varepsilon+\zeta~r^2)~\sin f~\cos f}{r^2+(\varepsilon+\zeta r^2)~\sin^2f}~v^2-\frac{1}{r}\left(\frac{r^2-(\varepsilon+\zeta~r^2)~\sin^2f}{r^2+(\varepsilon+\zeta~r^2)~\sin^2f}\right)~v\nonumber\\
&&+~\frac{n^2(1+\frac{\zeta}{\varepsilon}~r^2)~\sin f~\cos f}{r^2+(\varepsilon+\zeta~r^2)~\sin^2f}
\end{eqnarray}
Let us write down (\ref{41}), (\ref{42}) in iterative expression as
\begin{eqnarray}\label{44}
v_{i+1}=v_i+\frac{dr}{6}(k_1+2k_2+2k_3+k4)
\end{eqnarray}
and
\begin{eqnarray}\label{45}
f_{i+1}=f_i+\frac{dr}{6}(l_1+2l_2+2l_3+l_4)
\end{eqnarray}
with
\begin{eqnarray}\label{46}
l_1=v,~~~~~~~k_1=g(r,~f,~v)
\end{eqnarray}
\begin{eqnarray}\label{47}
l_2=v+\frac{dr}{2}k_1,~~~~~~~k_2=g(r+\frac{dr}{2},~f+\frac{dr}{2}l_1,~v+\frac{dr}{2}k_1)
\end{eqnarray}
\begin{eqnarray}\label{48}
l_3=v+\frac{dr}{2}k_2,~~~~~~~k_3=g(r+\frac{dr}{2},~f+\frac{dr}{2}l_2,~v+\frac{dr}{2}k_2)
\end{eqnarray}
\begin{eqnarray}\label{49}
l_4=v+dr~k_3,~~~~~~~k_4=g(r+dr,~f+dr~l_3,~v+dr~k_3)
\end{eqnarray}
Eq.(\ref{44}) is numerical solution of $df/dr$. 

We obtain energy per unit length, $\mu$, which can be derived from Lagrangian, $\mathcal{L}_2$, in eq.(\ref{28}). Replace $r$ with $\overline{r}=qr$ ($q$ is scale factor, a number) then energy per unit length can be written as
\begin{equation}\label{50}
\mu = 2\pi\int_0^{\infty} \Bigg[\Big(\frac{df}{d\bar{r}}\Big)^2+n^2\Big(\frac{1}{\bar{r}^2}+\frac{\zeta}{\varepsilon~q^2}\Big)\sin^2f+\Big(\frac{df}{d\bar{r}}\Big)^2\Big(\zeta+\frac{\varepsilon~q^2}{\bar{r}^2}\Big)~\sin^2f\Bigg]\bar{r}~d\bar{r}
\end{equation}
Eq.(\ref{50}) can be written as
\begin{equation}\label{51}
\mu = 2\pi\int_0^{\infty} \Bigg[\Big(\bar{r}+\zeta~\bar{r}~\sin^2f+\frac{\varepsilon~q^2}{\bar{r}}~\sin^2f\Big)\Big(\frac{df}{d\bar{r}}\Big)^2+n^2\Big(\frac{1}{\bar{r}}+\frac{\zeta~\bar{r}}{\varepsilon~q^2}\Big)~\sin^2 f\Bigg]~d\bar{r}
\end{equation}
From eq.(\ref{51}) we are able to define 
\begin{equation}\label{52}
\eta\equiv\Big(\bar{r}+\zeta~\bar{r}~\sin^2f+\frac{\varepsilon~q^2}{\bar{r}}\sin^2f\Big)\Big(\frac{df}{d\bar{r}}\Big)^2+n^2\Big(\frac{1}{\bar{r}}+\frac{\zeta~\bar{r}}{\varepsilon~q^2}\Big)\sin^2 f
\end{equation}
So, eq.(\ref{52}) can be rewritten as
\begin{equation}\label{53}
\mu=2\pi\int_0^{\infty}\eta~d\bar{r}
\end{equation}
From chain rule, we have the relation that
\begin{equation}\label{54}
\frac{\partial \mu}{\partial q}=\frac{\partial\mu}{\partial \bar{r}}~\frac{\partial\bar{r}}{\partial q}\longrightarrow\frac{\partial\eta}{\partial q}=\frac{\partial\eta}{\partial\bar{r}}~\frac{\partial\bar{r}}{\partial q}
\end{equation}
Refer to (\ref{53}) and (\ref{54}), we can derive relation as below
\begin{eqnarray}\label{55}
\frac{\partial\mu}{\partial q} 
&=&2\pi~\frac{\partial}{\partial q}\int_0^\infty\eta~d\overline{r}
=2\pi\int_0^{\infty}\frac{\partial\eta}{\partial q}~d\bar{r}
=2\pi\int_0^{\infty}\frac{\partial\eta}{\partial\bar{r}}~\frac{\partial\bar{r}}{\partial q}~d\bar{r}
=2\pi\int_0^{\infty}\frac{\partial\eta}{\partial\bar{r}}~r~d\bar{r}\nonumber\\
&=&\frac{2\pi}{q}\int_0^{\infty}\frac{\partial\eta}{\partial\bar{r}}~\bar{r}~d\bar{r}
\end{eqnarray}
(Integration and differentiation operations are interchangeable, if the function is continuous.) In order to evaluate the right hand side of eq.(\ref{55}), we use identity relation 
\begin{equation}\label{56}
\frac{\partial}{\partial\bar{r}}\big(\eta~\bar{r}\big)
=\frac{\partial\eta}{\partial\bar{r}}~\bar{r}+\eta~~~\longrightarrow~~~\frac{\partial\eta}{\partial\bar{r}}~\bar{r}=\frac{\partial}{\partial\bar{r}}\big(\eta~\bar{r}\big)-\eta
\end{equation}
Then, we obtain
\begin{eqnarray}\label{57}
\frac{\partial\mu}{\partial q} 
=\frac{2\pi}{q}\int_0^\infty\frac{d\eta}{d\overline{r}}~\overline{r}~d\overline{r}
=\frac{2\pi}{q}\int_0^\infty\left[\frac{\partial}{\partial\overline{r}}(\eta~\overline{r})-\eta\right]~d\overline{r}
=\frac{2\pi}{q}~\eta~\left.\overline{r}\right|_0^\infty-\frac{2\pi}{q}\int_0^\infty\eta~d\overline{r}
\end{eqnarray}
Substitute (\ref{52}), (\ref{53}) into eq.(\ref{57}), we obtain
\begin{eqnarray}\label{58}
\frac{\partial\mu}{\partial q}\nonumber
&=&\frac{2\pi}{q}\Bigg[\Big(\overline{r}^2+\zeta~\overline{r}^2~\sin^2f+\varepsilon~q^2~\sin^2f\Big)\Big(\frac{df}{d\bar{r}}\Big)^2+n^2\Big(1+\frac{\zeta~\bar{r}^2}{\varepsilon~q^2}\Big)\sin^2 f\Bigg]_{0}^{\infty}\nonumber\\
&&-~\frac{1}{q}~\mu
\end{eqnarray}
Necessary condition for stability of vortex solution requires 
\begin{eqnarray}\label{59}
\left.\frac{\partial\mu}{\partial q}\right|_{q=1}=0
\end{eqnarray}
From (\ref{58}), (\ref{59}), we obtain
\begin{equation}\label{60}
0=\frac{2\pi}{q}\left[\left(\overline{r}^2+\zeta~\overline{r}^2~\sin^2f+\varepsilon~q^2~\sin^2f\right)\left(\frac{df}{d\overline{r}}\right)^2+n^2\left(1+\frac{\zeta~\overline{r}^2}{\varepsilon~q^2}\right)\sin^2f\right]_0^\infty-\frac{1}{q}\mu
\end{equation}
For $q=1$, it has consequence that $\overline{r}\rightarrow r$ and $\mu\rightarrow\mu_{\text{min}}$. We obtain
\begin{equation}\label{61}
\mu _{\text{min}}= 2\pi\Bigg[\Big(r^2+\zeta~r^2~\sin^2f+\varepsilon~\sin^2f\Big)\Big(\frac{df}{dr}\Big)^2+n^2\Big(1+\frac{\zeta~ r^2}{\varepsilon}\Big)\sin^2 f\Bigg]_{0}^{\infty}
\end{equation}
where $\mu_{\text{min}}$ is minimum energy per unit length which fulfill the stability requirements. Relation between $\zeta$ parameter and minimum energy per unit length, $\mu$, is shown as \textbf{Figure 1} below:
\begin{center}
\includegraphics[scale=1]{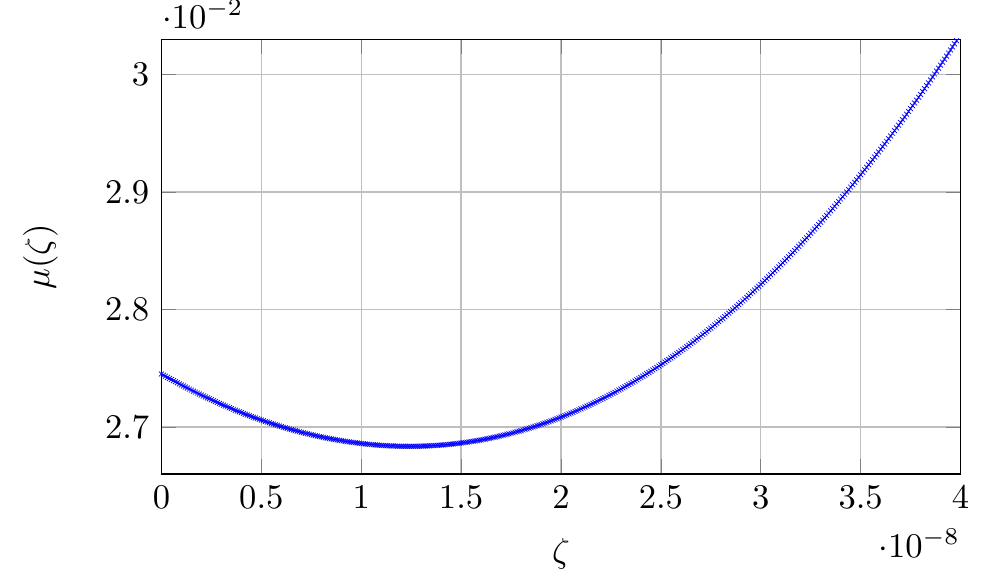}\\
\textbf{Figure 1}\\
Result for numerical solution of (\ref{61}): $\zeta$ versus $\mu$, with parameters $a=0.14,~dr=0.05,~\varepsilon=0.05,~n=1,~r_{\infty}=600$.
\end{center}
Here, we define $J$ as a point of $\zeta$ where $\mu$ is minimum. We find that the value of $J$ which is indicated by Matlab equal to 1.2400e-008. It means that the value of $J$ is equal to $1.24\times 10^{-8}$. We show both values, $J$ and $\mu_\text{min}$, in original form as shown by Mathlab in \textbf{Figure 2} below.
\begin{center}
\begin{tabular}{lcccc}
\hline\hline\\ $J$ &~&$\mu_{\text{min}}$~\\ 
[2.5ex] \hline\\ 1.2400e-008 && 0.0268  \\ 
\hline\hline
\multicolumn{2}{l}{}
\end{tabular}
\end{center}
\begin{center}
\textbf{Figure 2}\\
\text{$J$ and $\mu_{\text{min}}~\text{relation}$}
\end{center}

\section{Discussion}
Lagrangian density $\mathcal{L}_1$ (\ref{26}) is the Lagrangian of the nonlinear sigma model. The vortex solutions are scale-free: the energy per unit length, $\mu$, is independent of the width of the vortex. (We say that they the solutions are neutrally stable.)

Lagrangian density $\mathcal{L}_2$ (\ref{27}), (\ref{28}) are the Lagrangians with the Skyrme term added. The minimum energy per unit length occurs when the width of the untwisted vortex solution is infinite. So the vortex is unstable: it is energetically favourable for it to evaporate to infinity.

Lagrangian density $\mathcal{L}_3$ (\ref{35}) is the Lagrangian with a stabilising potential added. The minimum energy per unit length occurs at a finite value of the width of the untwisted vortex solution. So, the untwisted vortex with this particular width is stable. 

If instead of $\mathcal{L}_3$ we consider $\mathcal{L}_2$ and a twisted vortex solution (\ref{38}), then for a weak field, the equation of motion of the twisted $\mathcal{L}_2$ vortex is the same as the equation of motion of the untwisted $\mathcal{L}_3$ vortex. We expect the twisted $\mathcal{L}_2$ vortex solutions to also be stable at a finite value of the width. 

Necessary condition for stability of vortex solution has consequence that energy per unit length of vortex is minimum and scale-free. It means that vortex solutions are what is called neutrally stable to changes in scale. As scale factor change, the scale of the vortex changes, but the energy per unit length, does not. 

Let us discuss about unit of energy in more detail. The mass per unit length, $\mu$, of any cylindrically symmetric distribution of matter is usually quoted as a dimensionless quantity, meaning $G~\mu/c^2$ where $G$ is Newton's gravitational constant and $c$ is the speed of light. (In relativity it is conventional to use "geometrized" units, in which $c=G=1$, so the mass per length, $\mu$, the energy per unit length, $\mu c^2$, and the dimensionless quantity, $G\mu/c^2$, are numerically the same.) So, $\mu c^2$ (now the physical energy per unit length) is normally quoted in units of $c^4/G$ i.e. $c^4/G=1.2\times 10^{44}~\text{kg~m/s}^2=1.2\times 10^{44}~\text{J/m}=7.6\times 10^{62}~\text{eV/m}$. 

We find graphically the value of $\zeta$ parameter and energy per unit length, $\mu$, as shown in Figure 1. Figure 2 shows numerically that the value of minimum energy per unit length is 0.0268. It means that vortex minimum energy per unit length is $0.0268\times 7.6\times 10^{62}~\text{eV/m}=20.37\times 10^{60}~\text{eV/m}$.

\section{Acknowledgment}
MH thank to Professor Malcolm Anderson for long patience and clear guidance. Thank also to Professor Eugen Simanek, Professor Edward Witten and Professor Wojtek Zakrzewski for fruitful discussions. Thank to Professor Yongmin Cho and Professor Pengming Zhang who withdraw my attention on topological objects in two-component Bose-Einstein condensates, which is static version of baby Skyrmion cosmic string. Mr Andri Husein for numerical works and fruitful discussions. Professor Muhaimin, Dr Irwandi Nurdin for kindly help. Department of Mathematics Universiti Brunei Darussalam, and Physics Research Centre LIPI for support and huge chances for doing this research. All kindly colleagues for their strong supports in various ways. Profound gratitude to beloved mother, Siti Ruchanah, and beloved Ika Nurlaila, for continuous praying and sincere love. Beloved Aliya Syauqina Hadi for purity and her naughtiness. This research is supported fully by Graduate Research Scholarship Universiti Brunei Darussalam (GRS UBD).

\end{document}